# Preliminary Exploration on the Low-Pressure Ar-O₂ Plasma Generated by Low-Frequency Alternating Current (AC) Power Supply

Niaz Wali[1, *], W.W. Xiao[1], Q. U. Din[2], N. U. Rehman[2], C.Y. Wang[3], J.T. Ma[1], W.J. Zhong[1], Q.W. Yang[4,5]

1.  Institute for Fusion Theory and Simulation, School of Physics, Zhejiang University, Hangzhou, 310058, China
2.  Plasma Research Laboratory, Department of Physics, COMSATS University, Islamabad 45550, Pakistan
3.  Southwestern Institute of Physics, P.O. Box 432, Chengdu, China
4.  Key Laboratory of Biomass Chemical Engineering of Ministry of Education, College of Chemical and Biological Engineering, Zhejiang University, 310027 Hangzhou, Zhejiang, China.
5.   Institute of Zhejiang University-Quzhou, 324000 Quzhou, Zhejiang, China

Email: niaz@zju.edu.cn

This study reports a low-frequency alternating current (AC) power supply as a novel approach for generating low-pressure Ar-O2 plasma, offering advantages in cost, compactness, and operational simplicity, which are crucial for both material science and biological applications. The effectiveness of low-frequency AC-generated plasma against traditional RF systems by examining key plasma parameters such as electron density, electron temperature, and electron energy distribution function (EEDF), are investigated. Experimental results revealed that AC power supply could effectively produce low pressure Ar-O₂ plasma with comparable properties to RF systems. Most notably, the AC-generated plasma achieved a significant reduction in bacterial growth, suggesting its potential as a more economical and flexible alternative for enhancing plasma-assisted applications in sterilization and material processing.

## 1. Introduction

The generation and the physics of low-pressure plasma, particularly when utilizing inert gases like argon (Ar) mixed with reactive gases —such as $O_2$, $CF_4$, and $N_2$, have been extensively studied and applied in diverse fields including, deposition, etching, decontamination, bacterial inactivation, and others various industrial applications [1–3]. The $n_e$, the $T_e$, and the EEDF are useful for knowing the critical quantities in plasma processing [4,5]. The operational discharge parameters, such as the discharge type, the working gas pressure, the gas composition, and concentration, are important to get the $n_e$, the $T_e$ and the EEDF [6–9]. In recent decades, numerous experimental and theoretical studies have examined the impact of oxygen and nitrogen admixture on argon discharge for various medical and industrial purposes. There is extensive literature available on the RF inductively [10–15] and capacitively coupled Ar-$O_2$ plasma in low-pressure conditions [16–22].

The choice of power generator has significant influence on the generation and characteristics of low-temperature plasma. Different types of power supply offer distinctive advantages and are chosen based on the required plasma properties. DC power supplies, for instance, are renowned for their simplicity, which makes them suitable for industrial applications, such as, metal sputtering and coatings. Nevertheless, due to constant voltage supply, DC power supplies can cause considerable deterioration of the electrodes, resulting in the need for more frequent maintenance and inconsistency in the plasma process [23,24]. Radiofrequency (RF) power generators have traditionally dominated the field due to their effectiveness in industrial and plasma-based sterilization. However, the intricate technology involved in RF systems leads to a significant manufacturing and operational costs. These systems require sophisticated engineering to effectively control power outputs and frequency ranges [25,26]. The use of low frequency alternating current (AC) power supplies presents a novel approach, in low-pressure conditions. Based on our comprehensive literature review, there is a lack of studies examining Ar-$O_2$ admixture in capacitively coupled configurations under low-pressure conditions driven by the low-frequency AC power supply. This method offers distinct advantages in terms of cost-effectiveness, compactness, portable, has a direct electrode connection, and ease of operation, which are important in both materials' science and biological applications.

This study aims to bridge the gap in existing research by comparing the effects of low-frequency AC generated plasma to those typically achieved with RF systems, particularly in Ar-$O_2$ capacitively coupled plasma at low-pressure conditions. The effectiveness of a low-frequency

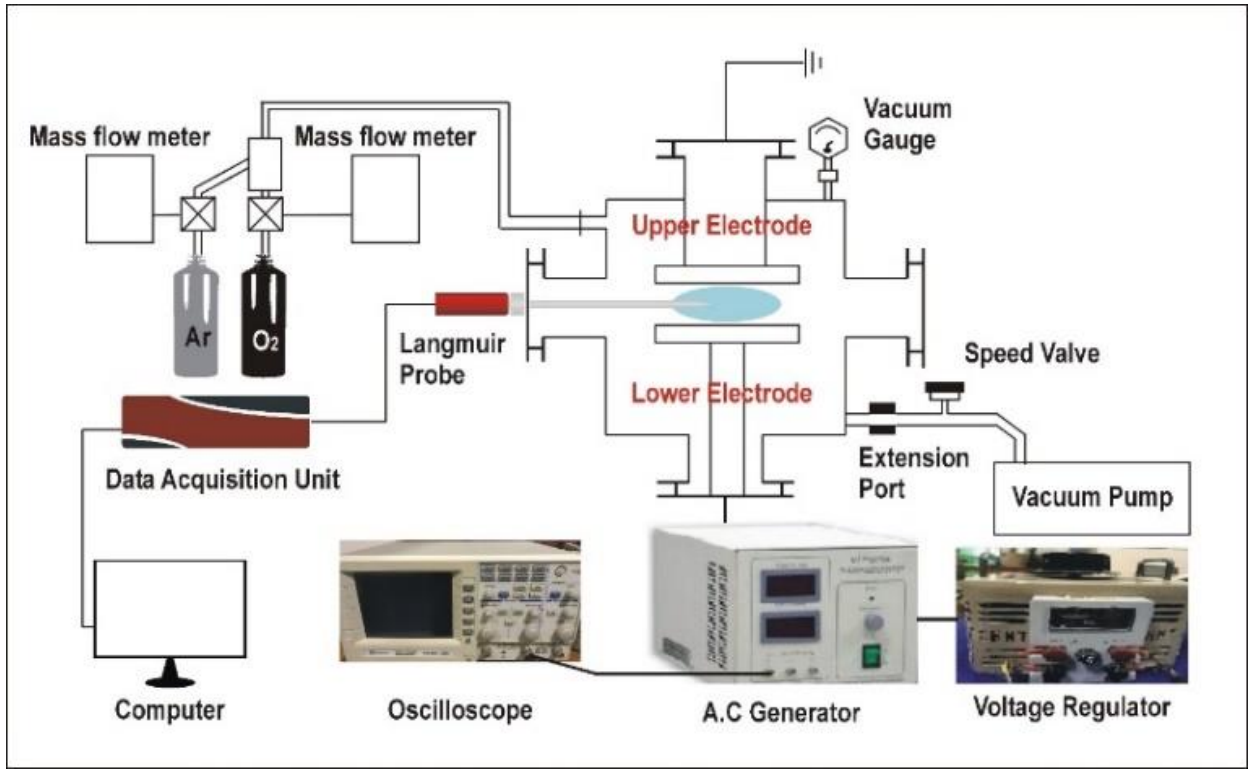

**Fig. 1.** Schematic diagram of the experimental setup with installed single Langmuir Probe.

AC-generated plasma is investigated against the inactivation of *Xanthomonas campestris pv. Vesicatoria (X. c. pv. vesicatoria)* bacteria. The reduction in bacterial growth following plasma treatment indicates that the Ar-O$_2$ plasma produced by a low-frequency AC power offers a valuable pathway to get the low-temperature plasma due to its affordability.

The paper is organized as follows: Section II describes the experimental setup and data analysis method obtained by a single Langmuir probe. The experimental results, including the $n_e$, the $T_e$, the plasma potential, the floating potential, the EEDF, and bacterial inactivation, are discussed in section III. The conclusion is presented in Section IV.

## 2. Material and Methods

### 2.1. Experimental Setup

The experiment is conducted in a stainless-steel vacuum chamber with an inner diameter of 39 cm and a height of 42.0 cm [19,27]. The Ar-O$_2$ plasma is generated in asymmetrical capacitively coupled electrodes installed with the low-temperature plasma experimental power supply, as shown in Fig. 1. The key diagnostic system is the single Langmuir probe in this experiment. Significantly, the AC power offers sinusoidal output voltage that ranges from 0 to 30 kV at 6 kHz.

The diameter of each electrode in the chamber is 14 cm, and a distance of 4.5 cm separates them. The AC power drives the lower electrode, and the rest of the chamber, including the upper electrode, is kept at zero potential, as shown in Fig. 1 in the red text.

Before feeding the mixture of gases, the vacuum chamber is pumped down to less than $3 \times 10^{-3}$ mbar for the discharge condition. A speed valve is used between the extension port and the rotary pump to isolate the plasma chamber from the pump. A Teledyne Hastings mass flowmeter controls the feeding gas flow, whereas the Pirani gauge records the working gas pressure. The filling gas pressure is increased from 0.5 mbar to 0.7 mbar by adjusting the gate valve, while the overall gas flow rate is fixed at 25 SCCM. The AC power at 6kHz is varied between 100 W and 900 W by employing the voltage regulator. Digital storage oscilloscope GDS-820S measures the operating frequency.

### 2.2. Analysis Method of a Single Langmuir Probe

The I-V characteristics are monitored by employing a single Langmuir probe (LP). Here, LP is made of tungsten wire with a radius of 0.195 mm, and just a length of 10 mm is inserted into bulk plasma. A 5 mm probe tip is exposed to plasma, and the remaining 5 mm is coated with ceramic. The probe tip is embedded into the reactor via a side window port to diagnose the bulk plasma, as shown in Fig. 1. The system has a computer-controlled power supply that can sweep the probe voltage from -20 V to +50 V while keeping a constant step voltage of 0.5 V.

The probe current is recorded during the experiment, while the voltage varies concerning the reference grounded electrode and the chamber wall. The probe tip is cleaned before each measurement by applying a bias voltage of 150 V using electron bombardment to avoid contamination affecting the probe I-V characteristics. The electron density, electron temperature, plasma potential ($V_p$), floating potential ($V_f$), and EEDF are automatically derived from the built-in Impedans Ltd automated Langmuir probe software [19]. Plasma potential is measured utilizing the second derivative zero crossing approach [28], whereas the $n_e$ and $T_e$ are obtained from the I-V characteristics curve using the probe current. The calculation equations are as follows:

$$\frac{1}{kT_e} = \frac{I(V_p)}{\int_{V_f}^{V_p} I(V)dV} \qquad (1)$$

and

$$n_e = \frac{I(V_p)}{A_p} \sqrt{\frac{2\pi m_e}{e^2 k_B T_e}}. \qquad (2)$$

Here, the equations (1) and (2) are generally deemed applicable for a Maxwellian distribution. The $k_B$ is the Boltzmann constant, $V_f$ and $V_p$ are the floating and plasma potential, respectively, $V$ is the probe biasing voltage concerning $V_p$, and the probe current is $I$. $A_p$ is the probe area. The $e$ and the $m_e$ represent the charge and mass of the electron, respectively.

The EEDF is determined by using the second derivative of the I-V characteristics, and the Druyvesteyn method is based on [29–31]:

$$\frac{d^2 I_e}{dV^2} = \frac{e^2 A_p}{4} \left( \frac{2e}{m_e V} \right)^{\frac{1}{2}} f_e(\varepsilon). \qquad (3)$$

Here, $f_e(\varepsilon)$, $\varepsilon$, $V$, $A_p$, $m_e$, and $e$ signifies the EEDF, the energy variable, the probe biasing voltage, the probe tip area, the electron mass, and the electron charge, respectively.

## 3. Experimental Results and Discussion

The experiment is carried out in a capacitively coupled plasma (CCP) chamber for a non-equilibrium $Ar - O_2$ plasma by the AC power supply. The current voltage (I-V) characteristics of Langmuir probes are used to determine the various plasma parameters. Figure 2 displays the I-V curve of a probe in argon and Ar-O₂ plasma, both at low-pressure of 0.5 mbar and operated at 400 watts of AC power. In both cases, the probe current increases with a rise in probe voltage, which is characteristic of a plasma response to an electric field.

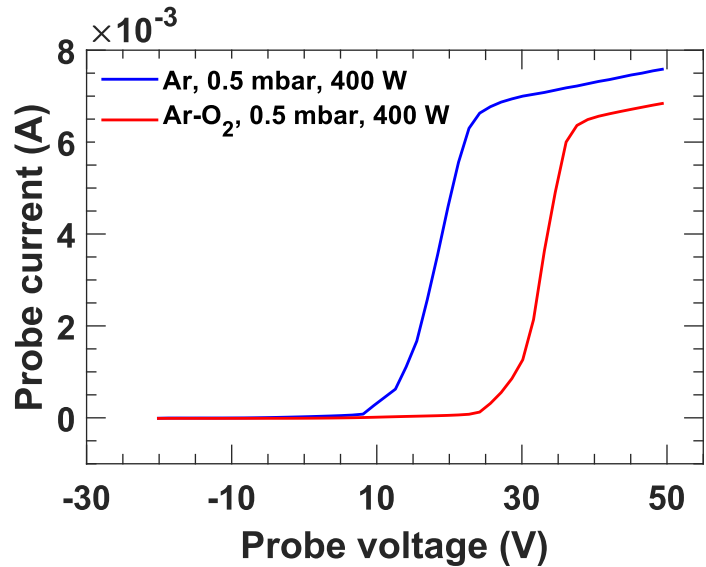

**Fig. 2.** The I-V characteristics of AC power supply generated Argon and Ar-O₂ (O₂, 4%) plasma at ~ 400 W and 0.5 mbar.

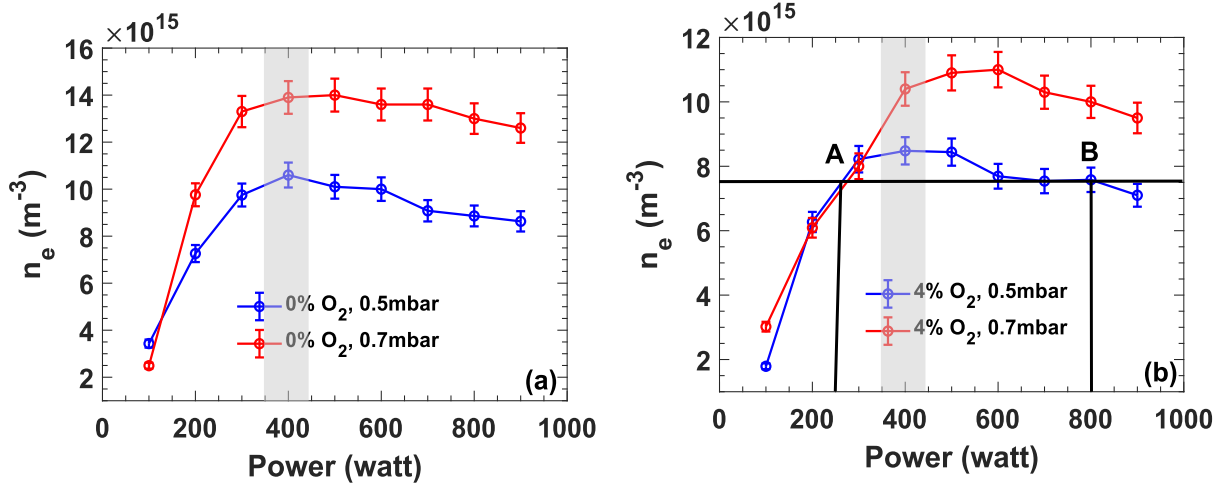

**Fig. 3.** (a) represents the evolution of the $n_e$ with different AC power supply of Argon plasma and (b) the evolution of the $n_e$ with different AC power supply of $Ar - O_2$ plasma, at 0.5 mbar and 0.7 mbar, respectively. The grey bars show a plasma density peak when the AC power is at ~ 400 W and 0.5 mbar.

It is clear from the figure that the current rise is more gradual and occurs at a higher threshold voltage in Ar-O2 compared to the pure argon plasma. This behavior is typically due to the presence of oxygen, which is an electronegative gas. $O_2$ molecules have greater electron affinity compared to argon, and have a greater tendency to capture electrons, leading to lower overall electron density [19]. This results in higher attachment rates, which affects the I-V characteristics and the zero-crossing point.

### 3.1. Plasma Electron Density

The density evolution of the Argon plasma and $Ar - O_2$ mixture plasma is shown in Figs. 3 (a) and 3 (b) as a function of applied power at different pressures. A nonlinear phenomenon of the $n_e$ change with AC power rising is observed in Fig. 3. The trend of the density changes is similar at different gas pressures both at 0% (pure argon) and 4% $O_2$ contents. A peak of the plasma density exists when the AC power is about 400 W, as shown in Fig. 3 by the grey bar.

The electron density $n_e$ can be altered with the change in gas pressure and oxygen concentrations as a function of the AC power. The increasing trend of the plasma density was noted with the input power increases from 100W to 300W, as shown in Fig. 3. It indicates that when the AC power increases, the electrons gain more energy due to the increased available electrical energy, which can create more ionizations to raise the electron density. When the AC power exceeds 400 W, the density decays with the increased applied power, as shown in Fig. 3. The slight decrease or saturation in $n_e$ at high power could be attributed to energy redistribution among various processes. This redistribution occurs possibly due to a change in the balance of electron production and loss mechanism, leading to a saturation or decrease in the $n_e$.

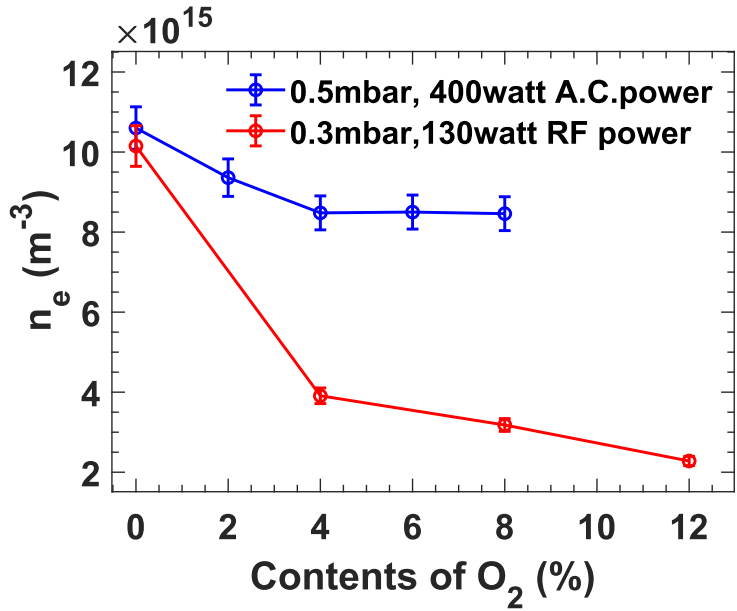

**Fig. 4.** The evolution of the $n_e$ with different $O_2$ contents at fixed AC power ~400 W and fixed gas pressure ~ 0.5 mbar, and the evolution of the $n_e$ with different $O_2$ contents at fixed RF power~130 W and fixed gas pressure ~ 0.3 mbar. Similar trends of density changes are observed for both types of discharges. The RF curve is deduced from the Fig 4 (c) of Ref [19].

The existence of a plasma density peak may be valuable to get needed densities in the low-temperature plasma because one can save the AC power based on the experimental goals. For example, if we hope to get the plasma density of the $Ar - O_2$ plasma is about $7 \times 10^{15} \ m^{-3}$, one can set the AC power is about 250 W at the "A" point or about 800 W at the "B" point, as shown in Fig. 3 by the black lines. The AC power at "A" point is more economically valuable to get the same plasma density. The plasma behavior at about 400 W of AC power is primarily studied because of the density peak.

Figure 4 shows the relationship between the $Ar - O_2$ plasma density and the $O_2$ contents at 400 W of the AC power at 0.5 mbar. With the fixed AC power (400 W) and the fixed gas pressure (0.5 mbar), the trend of the $n_e$ is decayed with $O_2$ contents increase, as shown in Fig. 4. The declining trend of the $Ar - O_2$ plasma density has been reported in the Ref. [19], which studied the evolution of the $n_e$ with different $O_2$ contents at fixed RF power (130 W) and fixed gas pressure (0.3 mbar), as shown in Fig. 4.

A comparison of different discharge types between the AC powered and the RF powered can be easily obtained based on the experimental results, as shown in Fig. 4. The RF curve, taken from the Ref. [19] where the dependence was observed at slightly different pressure, is shown here to compare the trend with the addition $O_2$ contents derived in the current work with other study. Thus, the AC power discharge is worth being deeply studied. In our work the preliminary study of the low-pressure Ar-$O_2$ plasma generated by AC power supply was performed.

### 3.2. Electron Temperature

The $T_e$ is another significant parameter in the Ar-$O_2$ plasma. Using the similar study method as mentioned above, we compared the results of the evolution of the $T_e$ with different AC power and different gas pressure. The $T_e$ is measured at 0.5 mbar and 0.7 mbar as a function of the AC power rising, respectively, as shown in Fig. 5 (a). Here, the blue curve is the evolution of the $T_e$ at 0.5 mbar, and the red curve is the evolution of the $T_e$ at 0.7 mbar. Figure 5 (a) indicates a decreasing trend in $T_e$ with both applied power and filling gas pressure while $O_2$ contents are maintained fixed. This declining tendency with rising applied AC power is caused by an increase in $n_e$, which raises the frequency of electron-electron collisions relative to electron-neutral collisions and lowers the $T_e$ [27].

Similarly, the electron temperature decreases with increasing filling gas pressure at a fixed frequency and $O_2$ contents, as depicted in Fig. 5 (a). This decrease can be accredited to the higher collision rate caused by increased pressure, which leads to energy being utilized in collisional processes like excitation, dissociation, and ionization. On the other hand, the increase in the electron temperature with the addition of $O_2$ contents is shown by the blue curve in Fig. 5 (b). The addition of $O_2$ contents causes the generation of vibrational and rotational excited states via electron collisions, which compete with electron impact ionization. Consequently, ionization decreases as electrons within the plasma are reduced through dissociative attachments and recombination with oxygen atoms and $O_2$ molecules. As a result, electron density decreases, reducing electron-electron collisions and allowing for a higher value of electron temperature.

The evolution of the $T_e$ by the AC power is also compared with the previous result [19] by the RF generator. A good agreement is observed between this work and the earlier results by the RF plasma, although the discharge types are different. The comparison indicates that the different discharge types, the AC power supply, and the RF power supply can cause similar behaviors in the Ar-$O_2$ low-pressure plasma. However, AC-powered plasma has a bit higher $T_e$ than RF plasma, possibly due to the difference in the power deposition mechanism. In AC, power is directly deposited into the plasma by the electrodes, while in RF, power is transmitted to the plasma through the sheath, resulting in reduced energy transfer to the electrons and, consequently, lower $T_e$ than AC-powered discharge.

### 3.3. Electron Energy Distribution Function (EEDF)

To understand the dynamic behaviors of the Ar-$O_2$ plasma, the Electron Energy Distribution Function (EEDF) has also been studied, even though we have compared the $n_e$ and the $T_e$ in different discharge types. The EEDF of the low-temperature plasma describes both the heating process and the numerous collisional processes [19,32,33]. Figure 6 (a) shows the variation of EEDF in different gas pressures while fixed AC power and oxygen concentration. Figure 6 (b) shows the variation of EEDF in different AC power while at fixed oxygen concentration and fixed pressure. The graph clearly shows the low-energy and high-energy electrons at any specific power and pressure. Generally, the peak at lower energies corresponds to the low-energy electrons, while the high-energy electrons correspond to the peak at higher energies. *Notes:* according to the definition, the energy distribution of low-energy electrons is usually in the range of several electron volts. These low-energy electrons are primarily responsible for the plasma chemical reactions and play a crucial role in plasma sustaining. High-energy electrons typically have energies ranging from ten to a few tens of eV [4,5,20]. They are responsible for plasma heating, ionization, and excitation of the gas or mixture of gases.

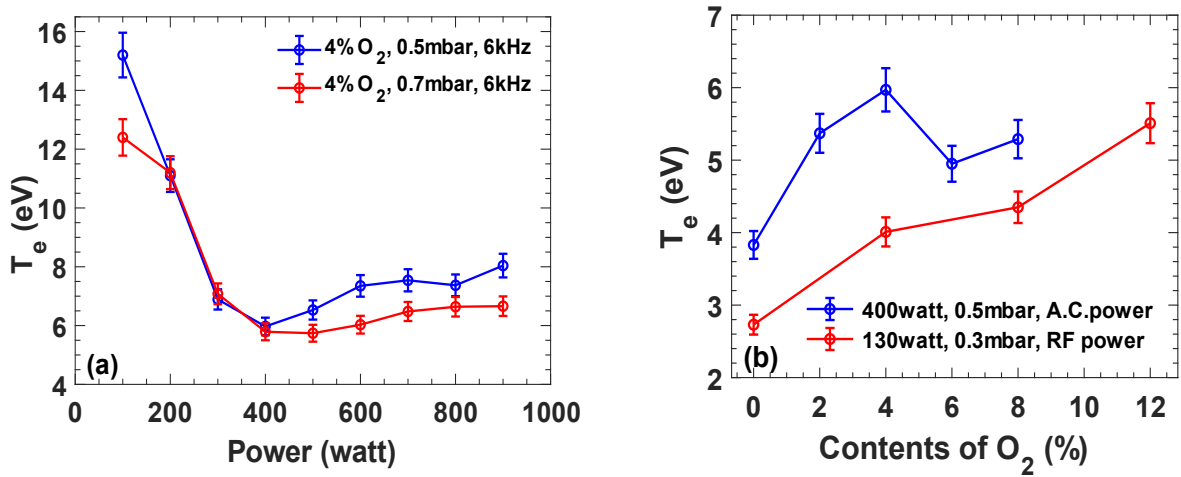

**Fig. 5.** (a) represents the evolution of the  $T_e$ with different AC power at 0.5 mbar and 0.7 mbar, respectively. (b) a comparison of the $T_e$ in different discharge types by the AC power supply and the RF power supply. The RF curve is deduced from the Fig 5 (c) of Ref [19].

The trend in Fig. 6 (a) indicates a reduction in the low-energy (particularly ≤ 3 eV) electrons population with a rise in gas pressure at constant applied power, suggests that initial the energy transfer to the gas molecules via elastic collision without causing ionization. Concurrently, as pressure increases, electrons may absorb energy through inelastic collisions which enhances the excitations and ionizations processes, contributing to the increased electrons population at energies beyond 3 eV. Lee et al., discussed the shift toward a bi-Maxwellian distribution and an increase in low-energy electrons with gas pressure in oxygen and in argon plasma [4]. A similar study has been documented in Ref. [34]. The non-Maxwellian shape is typical of the non-thermal plasmas such as used for material processing and sterilization, where electron energy is often dictated by a balance between electric field acceleration and collisional cooling processes.

Figure 6 (b) shows the EEDF at two different applied powers in Ar-O$_2$ plasma at fixed gas pressure. At the low input power of ~200 W, there is a higher concentration of high-energy electrons, while the high input power results in larger population of low-energy electrons, showing more efficient energy transfer to the plasma at ~600 W. This also might be possible that at high power, more electrons utilize their energy in inelastic collisions (exciting or ionizing gas atoms or molecules), leading to a reduction in the high-energy tail of the EEDF [35]. A pivotal energy level is identified by the precise crossover point at which the populations equalize for both the input power. Below this point, low-energy electron density increases with higher power, and above it decreases. This has an impact on the dynamics of chemical reactions in the plasma, where high-energy electrons drive ionization and excitation processes, and low-energy electrons facilitate adsorption and surface reactions.

These characteristics of the Ar-O$_2$ plasma by the AC power are similar to the case of the RF capacitively coupled plasma [19]. Experimental results suggest that this preliminary experiment on the low-pressure Ar-O$_2$ plasma generated by low-frequency AC power is worth studying and expanding on in the future.

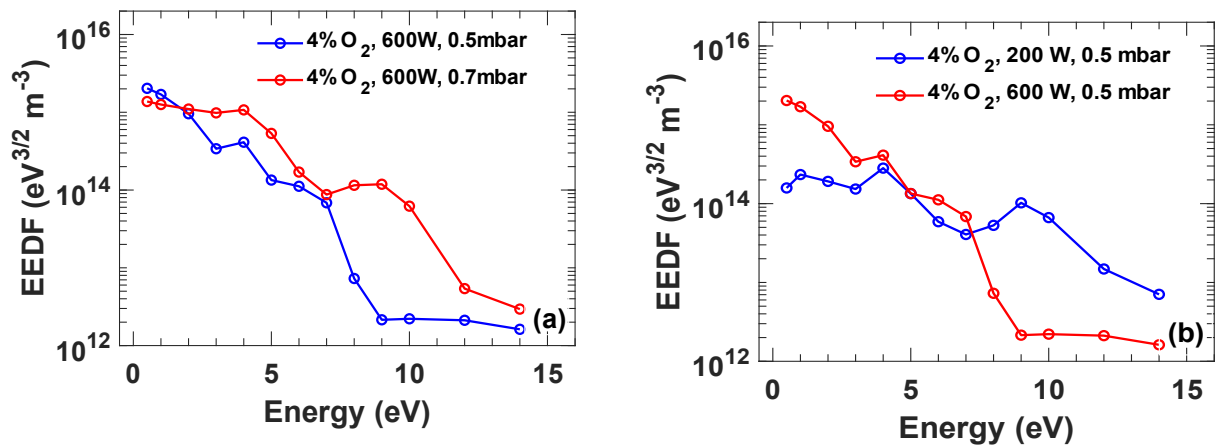

### 3.4. Bacterial Inactivation by AC Plasma treatment

Although causing fewer diseases and relatively less economic damage than fungi or viruses, plant pathogenic bacteria negatively impact the economic condition in many agricultural countries [36]. This study aims to evaluate the efficiency of low-frequency AC power supply generated plasma at low-pressure conditions for sterilizing pathogenic bacteria, a *Xanthomonas campestris pv. Vesicatoria (X. c. pv. vesicatoria)*, which is recognized for causing bacterial spot disease in plants, particularly affecting tomatoes, was treated. A bacterial suspension with a specific concentration of *X. c.* pv. *Vesicatoria,* roughly containing $10^6$ CFU/ml (CFU, colony forming units), was prepared and spread across the Petri dish. The prepared sample was carefully placed on the lower electrode. Subsequently, Ar-$O_2$ (4% $O_2$) was applied at a frequency of 6 kHz, maintaining a pressure of 0.5 mbar, and utilizing 400 W AC power for the plasma treatment process. Photographs of untreated and treated samples of *X. c.* pv. *vesicatoria* are shown in Figs. 7(a), (b), (c) and 7(d), respectively. Following the Ar-$O_2$ plasma treatment, the samples are kept in the incubator for 72 hours under 28 C$^o$ to observe the bacterial growth of untreated and treated samples. Our findings revealed a significant reduction of almost 75% in bacterial load following plasma treatment for 180 seconds, as shown in Fig. 7 (e). This promising result underscores the potential application of low-temperature plasma driven by a low-frequency AC power supply at lower pressures as an effective means of controlling and mitigating the impact of plant pathogens, offering a sustainable and environmentally friendly approach to enhance crop health and yield.

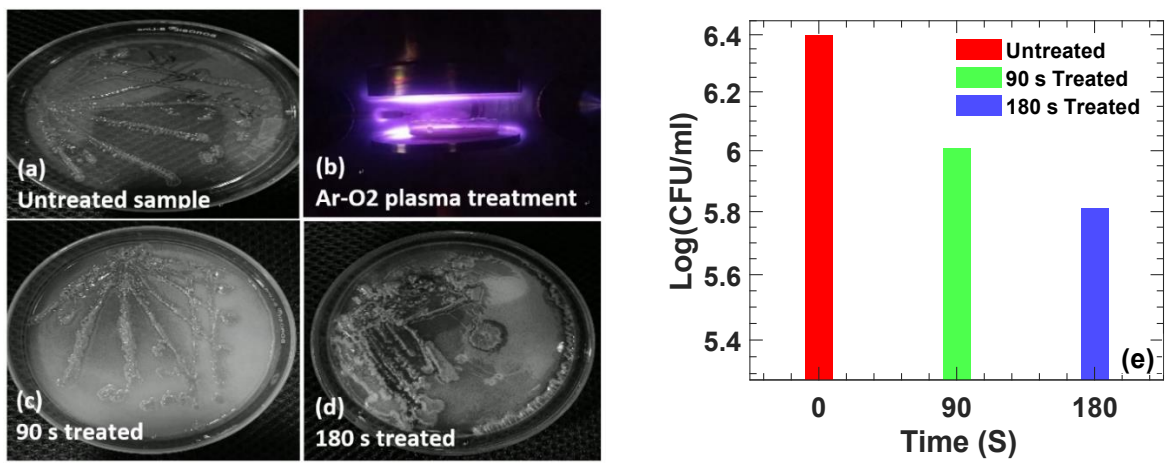

TABLE 1: A COMPARISON BETWEEN THE AC POWER SUPPLY AND THE R.F. SOURCE.

| Features | R.F. Source | AC power supply |
|---|---|---|
| Cost | High price [37] | Cheaper than R.F. source |
| Compatibility | Versatile [38] | Compatible with most devices |
| Power Output | High power output [39] | ~1000 VA of power output |
| Frequency Range | High frequency [38,39] | Low frequency up to 10 kHz |
| Portability | Inconveniently moveable, a matching network is required [38] | Portable |
| Noise | Require filtering or shielding [38] | Low distortion and noise |
| Safety | Potential exposure to radiation [40,41] | Safer |

## 4. CONCLUSIONS

The Ar $-$ $O_2$ plasma was generated by AC power with 6 kHz in low gas pressure conditions in the plasma experimental power supply for the first time. The plasma density peak suggests a useful application of this discharge type by AC power. The plasma behavior at about 400 W was studied as this work's main point. The EEDF was studied to understand the dynamic behaviors of the Ar-$O_2$ plasma in the AC power conditions. The non-Maxwellian shape of the EEDF in the Ar-$O_2$ plasma, characterized by the high EEDF tail, smoothed out with increasing applied power. In contrast, the tail of the EEDF increases with the pressure rising, which is the main characteristic of the AC power condition.

This work indicates that AC power supply possesses advantages differing from the RF power supply, such as being portable, inexpensive, safer, and simpler. The inactivation of *X. c. pv. vesicatoria* has been reported with the application of Ar-$O_2$ plasma. A considerable suppression of 75% for 180 s plasma treatment was recorded at 6 kHz. This suggests that the simple and cost-effective AC power generators may have useful applications for surface material treatment, biocompatible materials, biomedicine, crop-yielding growth, and plasma-based sterilization.

**Acknowledgment**

The authors thank Mr. Kamal Hussain and Dr. Zakia Anjum for the experimental support. This work is supported by the National Natural Science Foundation of China (Grant No.11875234).

**DECLARATIONS**

**Conflict of interest** The authors do not have any relevant financial or non-financial interests to declare.

**DATA AVAILABILITY STATEMENT** This manuscript has no associated data. The data that support the findings of this study are available from the corresponding author upon reasonable request.